# Large-scale Interconnection Power System Model Sanity Check, Tuning and Validation for Frequency Response Study

Shutang You[1], Yilu Liu[1,2]

1. University of Tennessee, Knoxville, TN, USA
2. Oak Ridge National Laboratory, TN, USA
Email: syou3@utk.edu

*Abstract*— **The quality and accuracy of power system models is critical for simulation-based studies, especially for studying actual stability issues in large-scale systems. With the deployment of wide-area monitoring systems (WAMSs), the high-reporting-rate frequency measurement provides a trustworthy ground truth for validating models in frequency response studies. This paper documented an effort to check, tune, and validate the U.S. power system model based on a WAMS called FNET/GridEye. Four metrics are used to quantitatively compare the simulation results and the actual measurement, including frequency nadir, RoCoF, settling frequency and settling time. After tuning governor deadband and the governor ratio, the model frequency response shows significant improvement and matches well with the event measurement data. This work serves as an example for tuning and validating large-scale power system models.**

## I. INTRODUCTION

Ensuring reliability is an important goal in power grid planning and operation. As the penetration of renewable energy increases, power grid planners and operators hope to look ahead on power grid conditions to make better preparation for a high renewable future. Therefore, they increasingly rely on future grid models for developing better investment and operation plans of power grids. However, it is fairly challenging to obtain accurate power grid models due to modeling errors and parameter inaccuracy. The development of wide-area synchrophasor measurement technology, especially low-cost synchrophasor sensor technology, has facilitated real-time monitoring and enabled a variety of situational awareness applications to improve grid reliability. Based on this measurement data, many other applications have been developed to improve grid reliability and stability. These measurements can also serve as a basis for modeling current grids and study future grid scenarios and explore future stability challenges. Nevertheless, little literature discussed how to validate grid dynamic models for frequency response studies using synchrophasor measurements.

Frequency response is important for systems to ensure frequency stability after a grid event. As noticed by both academia and industry, some interconnection models that are widely used in the industry exhibit frequency responses that are much more optimistic than system actual performances. Such inaccuracy tends to conceal the potential risks of frequency response degradation associated with PV penetration. Therefore, building a credible baseline model is critical to the correctness of the end results and conclusions of the entire project.

In this paper, efforts have been made to validate or check the dynamic models of the three U.S. interconnection models. By comparing simulated frequency response against multiple frequency events recorded by FNET/GridEye, model parameters can be tuned so that simulations are able to show much improved consistence with actual measurements. The validated interconnection models are ready for future studies on grid stability.

## II. EI MODEL SANITY CHECK

It has been noticed that the EI multi-regional modeling working Group (MMWG) models do not reflect actual system frequency response accurately (as shown in Figure 1) [1, 2]. When used for grid frequency response studies, the original EI models with inaccurate representation of frequency response tend to be blind-sighted to potential risks. Therefore, the EI MMWG dynamic model was tuned and validated in previous projects using FNET/GridEye frequency measurement [3].

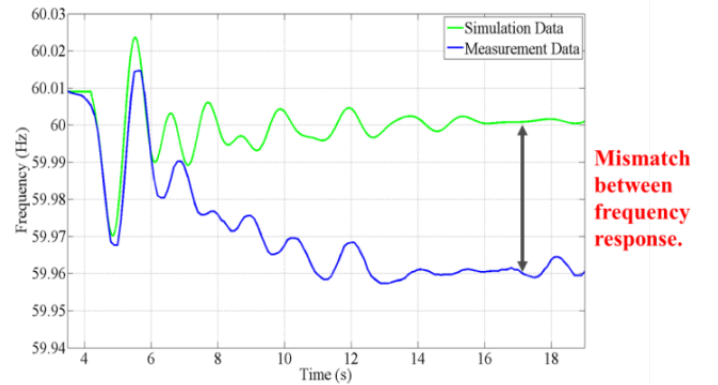

Figure 1. Frequency response mismatch between EI measurement and simulation

According to a previous study done by Oak Ridge National

This work was also supported by U.S. Department of Energy Solar Energy Technologies Office under award number 30844. This work made use of Engineering Research Center shared facilities supported by the Engineering Research Center Program of the National Science Foundation and the Department of Energy under NSF Award Number EEC-1041877 and the CURENT Industry Partnership Program.

Laboratory (ORNL), the fraction of generators that provide governor response (also referred to as governor ratio) and governor deadband are the most influential factors of interconnection-level frequency response [3]. Other factors, such as governor droop and generator inertia, also play a role. Governor ratio matters because it determines how many generation units are able to provide power support after a generation loss event. The unrealistically high governor ratio in the original EI MMWG model was the major reason why the measured settling frequency was much lower than the simulated value. Therefore, the original MMWG model's governor ratio was reduced in order to be consistent with the measured frequency response.

Governor deadband is another major factor. Since frequency can fluctuate within a range of tens of mHz under normal operation conditions, governor deadbands are implemented to prevent generators from excessive controls. This parameter was usually ignored in EI dynamic simulations. However, the EI system is such a large interconnection that its frequency deviation after a generation trip is within 50 mHz in many cases. In contrast, the typical governor deadband is 36 mHz [3]. Therefore, the impact of governor deadbands on frequency response should not be ignored for the EI system. In order to represent the impact of governor deadbands, a WECC-modified generic turbine governor model, WSIEG1 (as shown in Figure 2), was employed. This model can represent the dynamics of most thermal governors, and more importantly, capture the nonlinear characteristic of the governor deadband. To keep the consistency of governor parameters, original governor models, such as TGOV1, GAST, IEESGO, and IEEEG1, were converted to WSIEG1 through proper model logic conversion. An example of model logic conversion from TGOV1 (as shown in Figure 3) to WSIEG1 is listed in (1) to (5). Other conversion equations are documented in [3]. To model the overall governor deadband effect of the EI system, an average value of governor deadband was employed uniformly in the model.

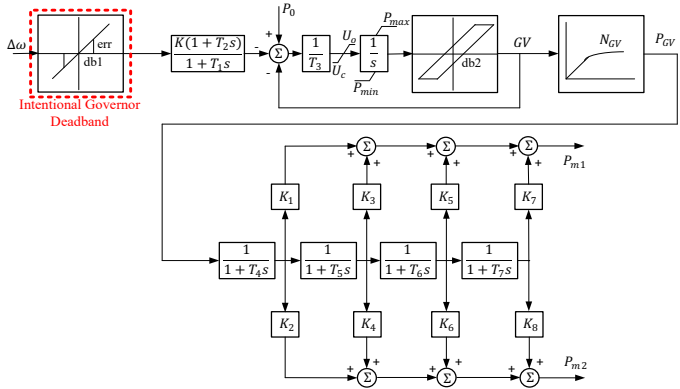

Figure 2. WSIEG1 block diagram

$$K = 1/R \quad (1)$$
$$T_3 = T_1 \quad (2)$$
$$K_1 = T_2/T_3 \quad (3)$$
$$K_3 = 1 - T_2/T_3 \quad (4)$$
$$T_5 = T_3 \quad (5)$$

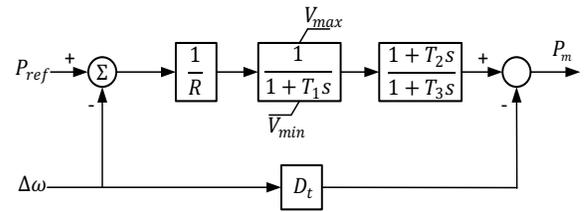

Figure 3. TGOV1 block diagram

Governor ratio, governor deadbands, and generator inertia, were adjusted in a coordinated manner to match actual FNET/GridEye frequency measurements. The FNET/GridEye frequency measurement accuracy is ±0.0005 Hz, which provides high accuracy for model tuning. After this tuning process, the EI MMWG model showed a much improved simulation accuracy compared with actual measurement.

In the following sections, multiple events recorded by FNET/GridEye will be used to perform a sanity check on the validated EI MMWG model. Two test events are given in details as examples and a summary of all the test cases is also presented afterwards.

### A. EI Case 1

The first test event is a 1,016 MW generation trip, whose detailed information can be found from the Nuclear Regulation Commission (NRC) website [1]. The frequency response comparison between simulated value and actual measurement of three different locations are shown in Figure 4. Clearly, the frequency nadir, rate of change of frequency (ROCOF), frequency settling time, and settling frequency all demonstrate good consistencies. Furthermore, as shown by Table 1, the .

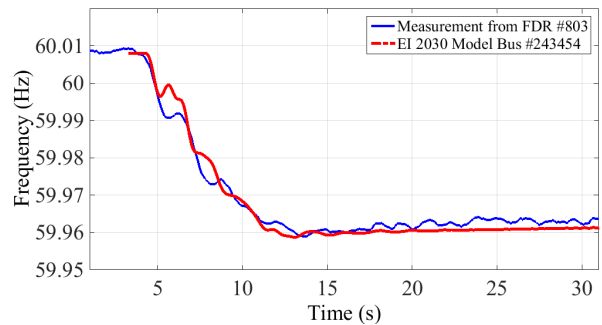

(a) Observation in Ohio

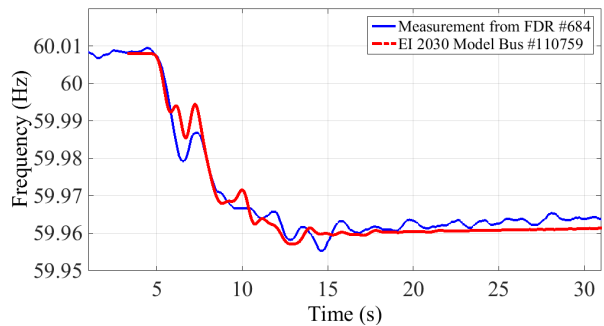

(b) Observation in Massachusetts

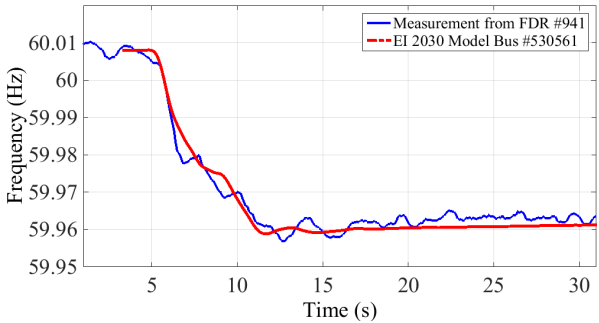

(c) Observation in Kansas

Figure 4. EI model sanity check results: EI case 1

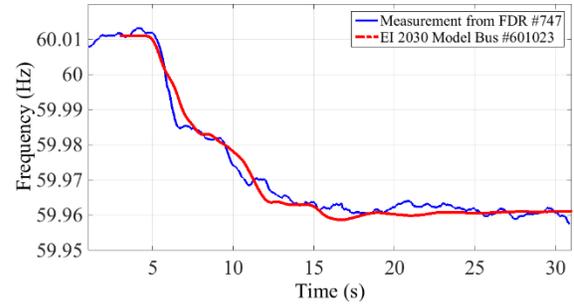

(c) Observation in Minnesota

Figure 5. EI model sanity check results: EI case 2

Table 1. EI model sanity check metrics for EI case 1

|  | FNET/GridEye Measurement | Simulated Value | Mismatch | Metric Success Value |
|---|---|---|---|---|
| Frequency Nadir (Hz) | 59.959 | 59.959 | 0.000 | 0.010 |
| Rate of Change of Frequency[2] (mHz/s) | 4.94 | 5.58 | 0.64 | 10 |
| Frequency Settling Time (s) | 9.9 | 9.2 | 0.7 | 3.0 |
| Settling Frequency (Hz) | 59.962 | 59.961 | 0.001 | 0.010 |

Table 2. EI model sanity check metrics for EI case 2

|  | FNET/GridEye Measurement | Simulated Value | Mismatch | Metric Success Value |
|---|---|---|---|---|
| Frequency Nadir (Hz) | 59.961 | 59.959 | 0.002 | 0.010 |
| Rate of Change of Frequency (mHz/s) | 4.39 | 4.83 | 0.44 | 10 |
| Frequency Settling Time (s) | 11.5 | 12.8 | 1.3 | 3.0 |
| Settling Frequency (Hz) | 59.960 | 59.963 | 0.003 | 0.010 |

Besides, Table 3 and Table 4 summarize the results of all four test cases used for the EI model sanity check. The metrics in Table 4 show that the validated EI base case meets the model evaluation metrics and it is ready for follow-on studies.

Table 3. Summary of EI test cases

|  | **Time (UTC)** | **Generation Trip Amount (MW)** |
|---|---|---|
| Case 1 | 2013/02/21 14:57:06 | 1,016 |
| Case 2 | 2013/05/28 19:07:54 | 974 |
| Case 3 | 2013/06/28 17:29:44 | 1,182 |
| Case 4 | 2013/ 03/ 12 18:51:50 | 921 |

## B. EI Case 2

The second test case is a 974 MW generation trip, whose detailed information is also provided by NRC[3]. Similar to case 1, frequency response comparisons at different locations are given in Figure 5 and the successfully-met metrics are presented in Table 2.

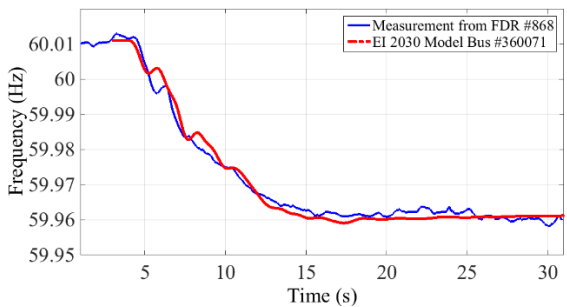

(a) Observation in Tennessee

Table 4. Summary of EI model sanity check metrics

| Mismatch | Frequency Nadir (Hz) | Rate of Change of Frequency (mHz/s) | Frequency Settling Time (s) | Settling Frequency (Hz) |
|---|---|---|---|---|
| Case 1 | 0.000 | 0.64 | 0.7 | 0.001 |
| Case 2 | 0.002 | 0.44 | 1.3 | 0.003 |
| Case 3 | 0.006 | 0.99 | 2.3 | 0.009 |
| Case 4 | 0.001 | 0.64 | 1.7 | 0.009 |
| Average Mismatch | 0.002 | 0.67 | 1.5 | 0.006 |
| Metric Success Value | 0.010 | 10.00 | 3.0 | 0.010 |

## C. Validation at Grid Edges for Case 1

Due to the large scale of the EI system, the need to check the model validation accuracy using FNET/GridEye measurements at grid edges was identified. Therefore, the accuracy of the validated EI model was checked at the EI grid edges using the FNET/GridEye measurement of the EI Case 1 event. Measurement locations at the EI edges are shown in Figure 6.

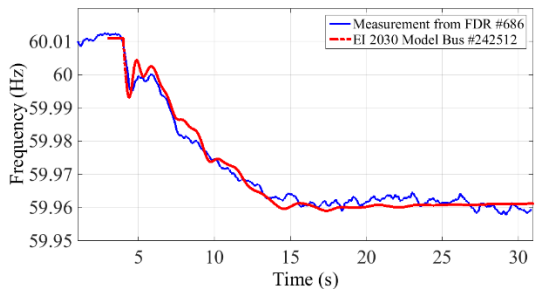

(b) Observation in Virginia

---

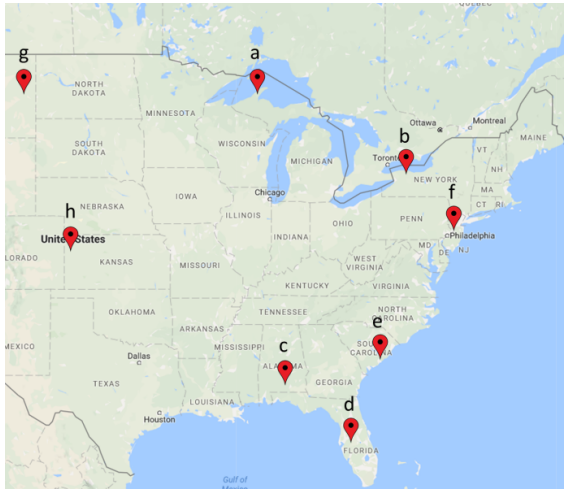

Figure 6. Measurement locations at the EI edges for model validation

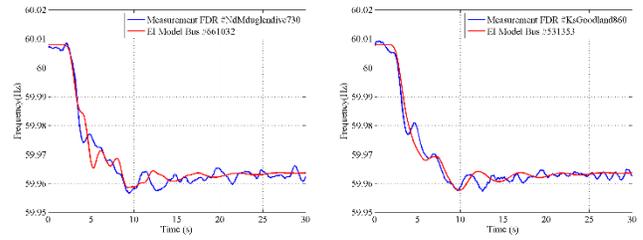

(g) Mdu glendive, ND    (h) Goodland KS

Figure 7. Model validation at the EI edges

## III.  ERCOT Model Validation

To improve the accuracy of the ERCOT model, governor deadbands were also incorporated into the ERCOT dynamic model. Similar to the work on the EI model, the WECC-modified generic turbine governor model was adopted in modeling the governor deadband in ERCOT. The original governor models, such as TGOV1, GAST, IEESGO, and IEEEG1, were translated to WSIEG1 through proper model logic conversion to keep the consistency of governor dynamics. The deadband settings are determined by ERCOT specifications [4].

The accuracy of the deadband-enabled ERCOT model was checked using the FNET/GridEye measurements. The case given here is a 390MW generation trip that occurred at 16:30:20 (UTC) on January 8, 2016. Figure 8 shows the measurements from the ERCOT system and the same event's simulation results from both models: one was tuned without deadband modeled while the other includes deadbands. Table 5 calculates the mismatches between measured and simulated values. It shows that the tuned model has high accuracy compared with measurement. In addition, the model with governor deadbands shows even smaller discrepancy in the frequency recovering period, as shown in Figure 8.

The simulation and measurement frequency response at these locations are shown in Figure 7, respectively. Despite some discrepancies of the low-magnitude oscillations, the simulation and measurement show good consistency in terms of overall frequency response.

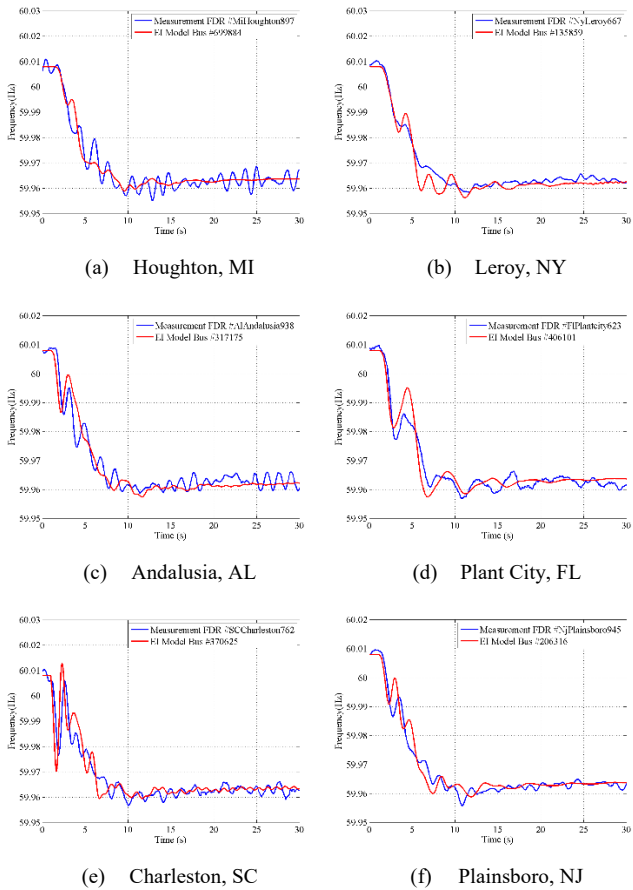

(a) Houghton, MI    (b) Leroy, NY

(c) Andalusia, AL    (d) Plant City, FL

(e) Charleston, SC    (f) Plainsboro, NJ

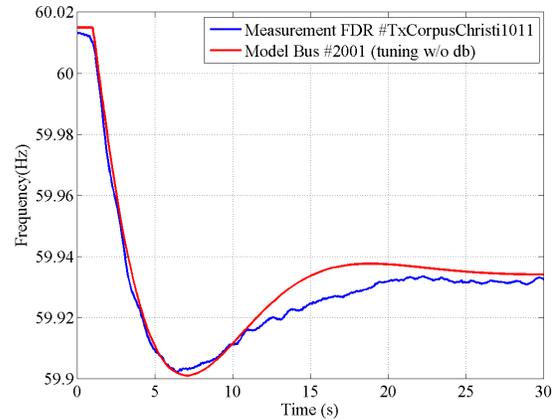

(a)  Model without deadband

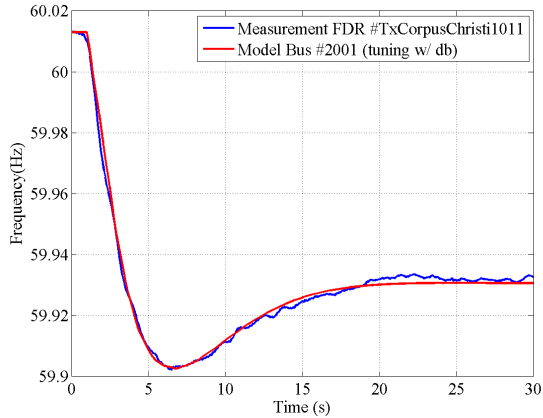

(b)  Model with deadband

Figure 8.  ERCOT model validation with and without governor deadband

Table 5.  ERCOT model validation metrics with and without governor deadband

| | FNET/GridEye Measurement | Simulation | | | | Metric Success Value |
|---|---|---|---|---|---|---|
| | | w/o db | Mismatch | w/ db | Mismatch | |
| Frequency Nadir (Hz) | 59.903 | 59.901 | 0.002 | 59.902 | 0.001 | 0.01 |
| Rate of Change of Frequency (mHz/s) | 37.6 | 37.7 | 0.1 | 37.7 | 0.1 | 10 |
| Frequency Settling Time (s) | 20.4 | 22.0 | 1.6 | 20.6 | 0.2 | 3 |
| Settling Frequency (Hz) | 59.930 | 59.935 | 0.005 | 59.931 | 0.001 | 0.01 |

Table 6 and Table 7 summarize the validation results of all the ERCOT test cases. Results demonstrate that all five events listed in Table 6 show high accuracy with actual measurement, which further proves the accuracy of the validated ERCOT dynamic model.

Table 6.  Summary of ERCOT test cases

| | Time | Generation Trip Amount (MW) |
|---|---|---|
| Case 1 | 2015/12/03 05:25:44 | 360 |
| Case 2 | 2015/12/21 21:19:53 | 320 |
| Case 3 | 2015/12/30 17:17:18 | 540 |
| Case 4 | 2016/01/08 16:30:20 | 390 |
| Case 5 | 2016/01/22 05:14:43 | 660 |

Table 7.  Summary of ERCOT model validation metrics

| Mismatch | Frequency Nadir (Hz) | Rate of Change of Frequency (mHz/s) | Frequency Settling Time (s) | Settling Frequency (Hz) |
|---|---|---|---|---|
| Case 1 | 0.003 | 7 | 2 | 0.003 |
| Case 2 | 0.003 | 2 | 0 | 0.004 |
| Case 3 | 0.01 | 6 | 3 | 0.006 |
| Case 4 | 0.001 | 0.1 | 0.2 | 0.001 |
| Case 5 | 0 | 5 | 1 | 0.009 |
| Average Mismatch | 0.004 | 4.0 | 1.5 | 0.005 |
| Metric Success Value | 0.01 | 10 | 3 | 0.01 |

## IV. CONCLUSIONS

This paper uses synchrophasor measurements from a wide-area monitoring system called FNET/GridEye to validate power grid models for frequency response studies. Four metrics of frequency response, including frequency nadir, RoCoF, frequency settling time, and settling frequency, are used to compare the model simulation results with the actual measurement. To reflect the system actual condition, the deadband is incorporated into governor models and the ratio of governors are tuned. The tuned models of U.S. EI and ERCOT systems show close match with the event frequency response in the FNET/GridEye database. These models can serve as base models for further studies on frequency response.